\documentclass[runningheads]{llncs}
\usepackage[T1]{fontenc}
\usepackage{graphicx}
\usepackage{booktabs}
\usepackage{algorithm}
\usepackage{algorithmic}
\usepackage{makecell}
\usepackage{enumitem}
\usepackage{multirow}
\usepackage{booktabs}
\usepackage{wrapfig}

\usepackage{subcaption}
\usepackage{amsmath}
\usepackage{dsfont}
\usepackage{amsfonts}   

\usepackage[misc]{ifsym}

\usepackage{mwe}

\begin{document}

\title{A Simple yet Effective Negative Sampling Plugin for Constructing Positive Sample Pairs in Implicit Collaborative Filtering}

\titlerunning{A Negative Sampling Plugin for Constructing Positive Sample Pairs}

\author{Jiayi Wu \and
	Zhengyu Wu \and
	Xunkai Li \and
	Rong-Hua Li\thanks{Correspondence to: Rong-Hua Li \textless lironghuabit@126.com\textgreater.} \and
	Guoren Wang}

\authorrunning{J. Wu et al.}

\institute{Beijing Institute of Technology, Beijing 100811, China\\
}

\maketitle              

\begin{abstract}
Most implicit collaborative filtering (CF) models are trained with negative sampling, where existing work designs sophisticated strategies for high-quality negatives while largely overlooking the exploration of positive samples. 
Although some denoising recommendation methods can be applied to implicit CF for denoising positive samples, they often sparsify positive supervision. 
Moreover, these approaches generally overlook user activity bias during training, leading to insufficient learning for inactive users. 
To address these issues, we propose a simple yet effective negative sampling plugin, PSP-NS, from the perspective of enhancing positive supervision signals. 
It builds a user–item bipartite graph with edge weights indicating interaction confidence inferred from global and local patterns, generates positive sample pairs via replication-based reweighting to strengthen positive signals, and adopts an activity-aware weighting scheme to effectively learn inactive users’ preferences. 
We provide theoretical insights from a margin-improvement perspective, explaining why PSP-NS tends to improve ranking quality (e.g., Precision@k/Recall@k), and conduct extensive experiments on four real-world datasets to demonstrate its superiority. 
For instance, PSP-NS boosts Recall@30 and Precision@30 by 32.11\% and 22.90\% on Yelp over the strongest baselines. 
PSP-NS can be integrated with various implicit CF recommenders or negative sampling methods to enhance their performance.

\end{abstract}

\section{Introduction}
Recommender systems were introduced to alleviate information overload~\cite{1} by helping users discover find content that matches their interests and needs amid the overwhelming information. By analyzing users’ historical behaviors, they provide personalized recommendations and improve user experience~\cite{2,9}.

Implicit feedback is widely used to model user preferences due to its non-intrusive nature~\cite{3} and ease of collection~\cite{4}. However, it provides only observed positives and lacks explicit negatives; thus, implicit CF typically adopts negative sampling~\cite{5}: observed interactions are treated as positives, while a subset of unobserved items are sampled as negatives~\cite{6}.
In this process, existing methods mainly focus on selecting high-quality negatives using signals such as item popularity~\cite{7} and user-behavior similarity~\cite{39,40,41,42}. 
However, they assume observed interactions are purely positive (e.g., clicks/purchases) and ignore negative feedback (e.g., skipping/hiding), which can be considered as false positives within positive sample pairs. Training with such positive sample pairs degrades model performance.

Some denoising recommendation methods~\cite{51,52,53,54,55}, though not designed for negative sampling, can be applied to negative sampling by leveraging training-time statistics (e.g., loss, gradients, variance) to denoise noisy positives and mitigate false positives. 
However, they suffer from two limitations: (1) \textbf{Positive signal sparsification}: their noisy/normal separation is often unreliable and not broadly applicable, leading to poor denoising accuracy.
Our experimental results (see Table 2) and \cite{55} confirm this phenomenon; 
consequently, mis-removing true positives sparsifies positive supervision and weakens training signals.
(2) \textbf{User activity bias}: they ignore user activity bias, which impacts the model's ability to learn the true preferences of inactive users.

To address these issues, we propose PSP-NS, a simple yet effective plugin for constructing Positive Sample Pairs in Negative Sampling. 
PSP-NS estimates interaction confidence from global and local signals, then constructs reliable positive pairs via replication-based reweighting to strengthen positive supervision.
This weakens noisy positives and introduces interactions consistent with global patterns, effectively enhancing the positive supervision signal and addressing Problem 1. 
Finally, PSP-NS applies activity-aware user weighting to adjust the focus of model training, addressing Problem 2.

\textbf{Our contributions.}
(1) \textit{\underline{New Perspective}}. 
We revisit negative-sampling-based implicit CF from a positive-signal construction perspective. 
Rather than crafting sophisticated negative samplers, we show that explicitly constructing high-quality positive training pairs within negative sampling can substantially improve learning.
(2) \textit{\underline{New Method}}. 
We propose PSP-NS, a simple yet effective negative sampling plugin that strengthens positive supervision by constructing confidence-weighted positive sample pairs from global patterns and local signals, while mitigating user activity bias via activity-aware user weighting. 
We further provide a margin-improvement insights to justify why these designs are effective.
(3) \textit{\underline{SOTA Performance}}. 
Extensive experiments demonstrate that PSP-NS consistently outperforms strong baselines and can be integrated into various implicit CF recommenders, working alongside different negative sampling strategies to further enhance model performance.

\begin{figure*}
	\centering
	\includegraphics[width=\linewidth]{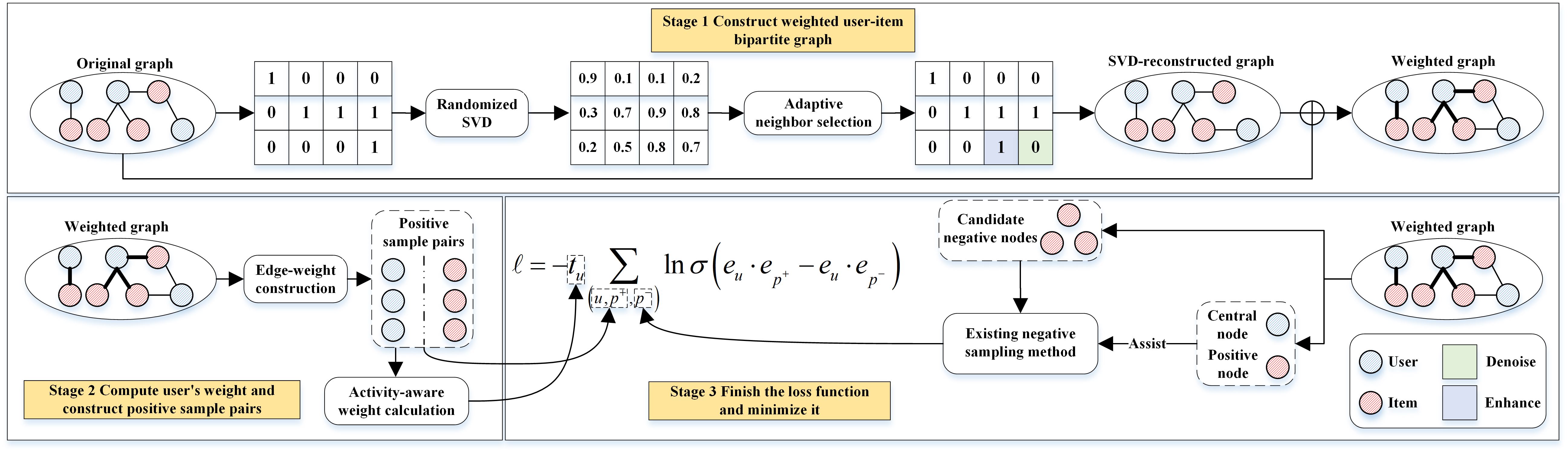}
	
	\caption{An overview of PSP-NS.}

\end{figure*}

\section{Related Work}

In negative sampling, most efforts focus on designing sampling distributions to mine high-quality negatives. 
RNS~\cite{56} uses a uniform distribution; PNS~\cite{7} uses a popularity-based distribution; DNS~\cite{11} and DNS(M,N)~\cite{48} adopt model-score-based sampling over candidates; IRGAN~\cite{57} learns a discriminator-guided distribution; SRNS~\cite{58} uses a prediction variance-aware distribution; MixGCF~\cite{39} uses a score-based distribution over mixup-augmented negatives; DENS~\cite{40} uses a sampling distribution based on the similarity of the relevant factor; and AHNS~\cite{49} uses a positive-score-conditioned distribution to sample negatives.

All the above negative sampling methods directly treat observed interactions as positives, overlooking false positives. 
Although not specifically designed for negative sampling, denoising recommenders can be applied to the negative sampling to alleviate this issue. 
R-CE and T-CE~\cite{51} reweight samples or truncate high-loss samples based on per-sample loss;
DeCA~\cite{52} exploits cross-model prediction consistency for positive denoising;
BOD~\cite{53} accumulates recent training signals in a parameter matrix and alternates optimization to downweight noisy interactions; DCF~\cite{54} uses the mean/variance of loss to filter noise while retaining hard samples; and PLD~\cite{55} performs resampling positive based on each user’s personalized loss distribution.

The existing denoising methods are almost designed based on observations from specific datasets, limiting generalizability.
They also heavily depend on model representations, causing unstable noise identification across training stages and thus low denoising accuracy that further sparsifies positive supervision signals. 
Motivated by this, we propose PSP-NS. Unlike prior denoising methods, PSP-NS mitigates noisy interactions by augmenting high-confidence positive pairs and further strengthens supervision by introducing new pairs consistent with global interaction patterns. 
Moreover, PSP-NS incorporates activity-aware user weighting to eliminate user activity bias. 
Experiments and theoretical insights validate its effectiveness in capturing true preferences and improving recommendation performance.

\section{Methodology}
In this section, we introduce PSP-NS. As shown in Figure~1, it has three stages: 
(1) We reconstruct $G$ via SVD to obtain $G_{\text{SVD}}$, and fuse $G_{\text{SVD}}$ with $G$ to build a weighted graph $\hat G$ that captures global patterns and local features.
(2) Using edge weights in $\hat G$ as confidence, we construct positive pairs via replication-based reweighting and compute activity-aware user weights.
(3) We then train the recommender by minimizing a negative sampling loss that combines these positive pairs and user weights with negatives from existing sampling methods.

\subsection{Construct the User-Item Bipartite Graph}
Given an implicit dataset $D = \left\{ {\left( {u,{p^ + }} \right) |u \in U,{p^ + } \in P} \right\}$, where $\left( {u,{p^ + }} \right)$ denotes an observed interaction, we construct a user–item bipartite graph $G=(V,E)$ with user and item nodes $V$ and interaction edges $E$. 
We then build the interaction matrix $A\in\{0,1\}^{|U|\times|P|}$, where $A(u,p)=1$ if $u$ interacts with $p$ and $0$ otherwise. To mitigate the popularity bias issue, we normalize $A$ as follows:
\begin{equation}
	\tilde A\left( {u,p} \right) = A\left( {u,p} \right)/\sqrt {rowD\left( u \right) * colD\left( p \right)} 
\end{equation}
where $rowD(u)$ and $colD(p)$ denote the degrees of user $u$ and item $p$, respectively.

Subsequently, we apply SVD to $\tilde A$, as prior work~\cite{30,43} shows that SVD can capture global interaction patterns by extracting dominant latent structures while suppressing noise. 
Formally, $\mathrm{SVD}(\tilde A,q)=M_q\Sigma_q N_q^{\top}$, where $M_q\in\mathbb{R}^{|U|\times q}$ and $N_q\in\mathbb{R}^{|P|\times q}$ are the user and item latent factor matrices, respectively, and $\Sigma_q\in\mathbb{R}^{q\times q}$ is a diagonal matrix whose entries are the singular values (factor strengths).
However, implicit feedback is typically sparse and large-scale, making full SVD difficult to converge and computationally expensive. 
Therefore, we adopt randomized SVD~\cite{27}, which performs SVD on a low-dimensional sketch of $\tilde A$ obtained via random projection and maps the results back to the original space. 
The formulas are as follows:
\begin{equation}
	Randomized\;\text{SVD}\left( {\tilde A,q} \right) = {\tilde M_q}{\tilde \Sigma _q}\tilde N_q^T\quad 
\end{equation}
\begin{equation}
	{\tilde A_{\text{SVD}}} = {\tilde M_q}{\tilde \Sigma _q}\tilde N_q^T
\end{equation}
To construct ${G_{\text{SVD}}}$ from ${\tilde A_{\text{SVD}}}$, we adopt an adaptive neighbor selection strategy. 
For each user $u$, we treat the $u$-th row of ${\tilde A_{\text{SVD}}}$ as a preference embedding over items and select the top-$K$ items as $u$’s neighbors. 
$K$ is set to each user’s historical interaction count, keeping user activity in ${G_{\text{SVD}}}$ consistent with $G$. Compared with a fixed $K$ (which requires dataset-specific tuning and may introduce noise or miss true preferences), this adaptive design avoids manual tuning and generalizes better across datasets. 
It also matches the intuition that users with more interactions provide richer signals, allowing us to identify more candidate items consistent with global interaction patterns, whereas less active users offer fewer signals and thus fewer such candidates~\cite{59}.

PSP-NS constructs ${G_{\text{SVD}}}$ via adaptive neighbor selection to add interactions consistent with global patterns while filtering noisy ones, but this process may discard some local signals. 
We therefore enhance ${G_{\text{SVD}}}$ with local interaction features from $G$. Nonetheless, this approach may reintroduce the noisy data that were previously removed.
To reduce its impact, we build a weighted bipartite graph $\hat G$, where edge weights downweight noisy interactions. 
For any $u\in U$ and $p\in P$, the edge weight in $\hat G$ is defined as:
\begin{equation}
	\hat W\left( {u,p} \right) = \left\{ {\begin{array}{*{20}{c}}
			{0,p \notin P_u^G\;and\;p \notin P_u^{{G_{\text{SVD}}}}}\\
			{s,p \in P_u^G\;and\;p \in P_u^{{G_{\text{SVD}}}}}\\
			{1,otherwise\quad \quad \quad \quad \quad\;{\kern 1pt} {\kern 1pt} }
	\end{array}} \right.
\end{equation}
where $P_u^G$ and $P_u^{G_{\text{SVD}}}$ are $u$’s interacted-item sets in $G$ and $G_{\text{SVD}}$, respectively.

\subsection{Construct Positive Sample Pairs and Compute User’s Weights}

\textit{Construct Positive Sample Pairs.}
In $\hat{G}$, the edge weight $\hat{W}(u,p)\in\{1,s\}$ reflects interaction confidence: $\hat{W}(u,p)=s$ indicates stronger global–local agreement (higher confidence), while $\hat{W}(u,p)=1$ is relatively lower-confidence.
To exploit this confidence signal, we design a simple yet effective strategy to construct the Positive Sample Pairs ($\text{PSP}$) via replication-based reweighting:
if $\hat{W}(u,p)=1$, we include $(u,p)$ once in $\text{PSP}$; if $\hat{W}(u,p)=s$, we include $(u,p)$ $s$ times.
This design is plug-and-play (no modification to the backbone) and, compared with merely multiplying the loss of a single triplet by $s$, is particularly effective under negative sampling:
replicating $(u,p)$ $s$ times often results in multiple independent samples of negatives across updates, yielding multiple constraints with different negatives and thus providing richer supervision (See~Table 3).

\textit{Compute User’s Weights.} 
In recommendation data, inactive users have far fewer interactions than active ones, making their preferences harder to learn. 
Inspired by the Inverse User Frequency principle~\cite{60} (originally used to down-weight popular items~\cite{60,61}), we extend the idea to the user dimension by up-weighting inactive users to reduce activity bias. 
Since user activity can be highly skewed, we adopt a simple nonlinear weighting function. For each $u\in U$, let $P_u^{\hat G}$ be the set of items interacted by $u$ in $\hat G$, and we compute user weight as follows:
\begin{equation}
	{t_u} = 1/{{\log \big( {a * \big| {P_u^{\hat G}} \big| + 1} \big)}}
\end{equation}
where $a$ denotes the sensitivity for calculating user weights, used to control the range of variation in the generated weight values. Adding 1 in the denominator prevents division by zero. As $\big| {P_u^{\hat G}} \big|$ increases, the user's weight decreases.

\subsection{Model Optimization}

In this section, based on the constructed $\text{PSP}$ and calculated user weights $t_u$, we minimize the following loss function to train the recommendation model.

\begin{equation}
	\ell  = \sum\limits_{\scriptstyle\left( {u,{p^ + }} \right) \in \text{PSP}\hfill,\scriptstyle{p^ - } \sim f\left( u \right)} { - {t_u}\ln \sigma \left( {{e_u^T}  {e_{{p^ + }}} - {e_u^T} {e_{{p^ - }}}} \right)} 
\end{equation}
where $\sigma \left(  \cdot  \right)$ denotes sigmoid function, and ${p^ - } \! \! \sim \! \! f\left( u \right)$ denotes negative samples from the negative sampling method. 

As shown in Algorithm~1, PSP-NS consists of three parts: (1) construct the weighted bipartite graph via randomized SVD and adaptive nearest neighbors (lines 1–4); (2) build positive sample pairs and compute user weights (lines 5–9); and (3) train the implicit CF model with negative sampling (lines 10–15).

\begin{algorithm}[t]
	\renewcommand{\algorithmicrequire}{\textbf{Input:}}
	\renewcommand{\algorithmicensure}{\textbf{Output:}}
	\caption{The training process with PSP-NS}
	\label{alg:1}
	\begin{algorithmic}[1]
		\REQUIRE Training set $D_{train}=\left\{ {\left( {u,{p^ + }} \right)} \right\}$, Recommendation model $Rec$, Number of singular values retained $q$, Sensitivity $a$ for calculating user weights, Edge weight $s$ in graph $\hat G$, Negative sampling method $f$.
		
		\STATE Construct the user-item interaction matrix $A$ and bipartite graph $G$ from $D_{train}$.
		\STATE Compute $\tilde A_{\text{SVD}}$ from $A$ via normalization and randomized SVD using (1-3).

		\STATE Construct $G_{\text{SVD}}$ from $\tilde A_{\text{SVD}}$ using the adaptive nearest neighbors method.
		\STATE Merge $G$ and $G_{\text{SVD}}$ and compute edge weights $\hat W$ via (4) to obtain $\hat G$.

		\STATE $\text{PSP} \leftarrow \emptyset $, $T \leftarrow dict()$.
		\FOR{each user $u \in U$}
		\STATE Compute $T[u]$ by (5).
		\STATE For each $(u,p)$ with $\hat W(u,p)=s$, add $(u,p)$ into PSP $s$ times.
		\ENDFOR
		\FOR{$t = 1,2,\dots,T$}
		\STATE Sample a mini-batch $\text{PSP}_{\text{batch}}$ from PSP.
		\STATE For each $(u,p^+) \in \text{PSP}_{\text{batch}}$, sample a negative item $p^- \sim f(u)$.
		\STATE $\ell  = \sum\limits_{} { - {t_u}\ln \sigma \left( {{e_u} \cdot {e_{{p^ + }}} - {e_u} \cdot {e_{{p^ - }}}} \right)}$.
		\STATE Update the parameters of the CF model $Rec$ by descending the gradients ${\nabla _\theta }\ell $
		\ENDFOR
	\end{algorithmic}
	
\end{algorithm}

\subsection{Theoretical Insights}
We provide theoretical insights into why PSP-NS is effective.
In particular, we show that (i) the positive set constructed by PSP-NS provides a stronger expected margin-improvement signal,
and (ii) activity-aware weighting amplifies the first-order margin gains for inactive users.

\noindent \textbf{Proposition 1.} \textit{For negative-sampling-based training with a smooth logistic loss (e.g., BPR), considering a single mini-batch update in a stable-training regime where the objective is locally $L$-smooth in the parameter region visited by training and the stochastic gradients have finite second moments (e.g., under standard $\ell_2$ regularization, finite embedding dimension, and bounded mini-batch size), PSP-NS can increase the expected pairwise margin, which is often reflected in improved ranking performance such as P@k/R@k in practice.}

\noindent\textit{Proof sketch.}
Let $s_\theta(u,p)=e_u^\top e_p$ and define the pairwise margin
$m = s_\theta(u,p^+) - s_\theta(u,p^-)$ for a triple $(u,p^+,p^-)$.
The BPR loss is $\ell(m)=-\log\sigma(m)$.
Consider a first-order gradient update $\theta^{+}=\theta-\eta\, \nabla_\theta \ell$ with a small effective step size $\eta$.
Let $m_{\theta}$ and $m_{\theta^{+}}$ denote the margins before and after the update, respectively.
By a first-order Taylor expansion of $m_\theta$ along $\theta^{+}-\theta=-\eta\nabla_\theta\ell$, we have
\begin{equation}
	m_{\theta^{+}}-m_{\theta}
	=
	-\eta\left\langle\nabla_\theta m_{\theta},\,\nabla_\theta\ell\right\rangle
	+
	r_\theta,
\end{equation}
where under local $L$-smoothness the remainder satisfies
$r_\theta \ge -\frac{L}{2}\|\theta^{+}-\theta\|^2 = -O(\eta^2)$.
Using $\nabla_\theta\ell=(\partial\ell/\partial m)\nabla_\theta m$ with
$\partial\ell/\partial m\!=\!-(1-\sigma(m))$, we obtain
\begin{equation}
	\mathbb{E}\!\left[m_{\theta^{+}}-m_{\theta}\right]
	\;\ge\;
	\eta\,\mathbb{E}\!\left[\left(1-\sigma(m_{\theta})\right)\left\|\nabla_\theta m_{\theta}\right\|^2\right]
	-
	O(\eta^2).
\end{equation}
Thus, the leading $O(\eta)$ term of the expected margin change is
$(1-\sigma(m_{\theta}))\|\nabla_\theta m_{\theta}\|^2$.
However, this surrogate improvement is only beneficial when the $p^+$ is indeed a true positive for $u$; otherwise the update can increase margins for false positives and bias the gradients.
Equivalently, it is governed by how often the training positives are true positives for $u$.

To this end, we quantify the quality of the constructed positives for user $u$ using two metrics: accuracy and coverage.
Let $S_u^{+}$ denote the constructed positives for $u$ and $D_u$ denote the ground-truth positives of $u$.
\begin{equation}
	\mathrm{Acc}(S_u^{+})=\mathbb{P}(p\in D_u \mid p\in S_u^{+}), \qquad
	\mathrm{Cov}(S_u^{+})=\mathbb{P}(p\in S_u^{+} \mid p\in D_u).
\end{equation}
$\mathrm{Acc}$ captures the true-positive rate of training positives, while $\mathrm{Cov}$ captures how many true positives are covered; larger values of both indicate more reliable positive signals.

PSP-NS is designed to increase $\mathrm{Acc}(S_u^{+})$ by upweighting (replicating) positives jointly supported by both the global SVD reconstruction and the observed local interactions, thereby increasing the proportion of high-confidence positives among training pairs.
It is also designed to increase $\mathrm{Cov}(S_u^{+})$ by injecting additional positives suggested by global interaction patterns via adaptive neighbor selection; while such injected pairs may still contain noise, emphasizing global-local agreement substantially enlarges the high-confidence subset, so the relative noise ratio in the training positives is reduced.
Together, these mechanisms mitigate positive-signal sparsification by providing more reliable and more complete supervision.
Therefore, compared to treating all observed interactions as positives in standard negative sampling or using denoising-only schemes, PSP-NS provides more reliable and more complete supervision, thereby strengthening the expected margin-optimization signal and tending to increase the expected pairwise margin; this empirically translates into improved P@k/R@k (as validated in Table~2).

\noindent \textbf{Proposition 2.}
\textit{Under the same stable-training assumptions as Proposition~1, activity-aware weighting amplifies the leading expected margin improvement for inactive users by a factor linear in $t_u$, thereby tending to improve their ranking quality (e.g., $P@k/R@k$) in practice.}

\noindent \textit{Proof sketch.}
Proposition~1 shows that for BPR training, a single mini-batch gradient descent step yields an expected margin change of the form
\begin{equation}
	\mathbb{E}[m^{+}-m]
	\;\ge\;
	\eta\,\mathbb{E}\!\left[(1-\sigma(m))\|\nabla m\|_2^2\right]
	\;-\;O(\eta^2).
	\label{eq:prop1_margin_template}
\end{equation}
With activity-aware weighting, the loss becomes $\ell_u(m)=-t_u\log\sigma(m)$, hence
$\frac{\partial \ell_u}{\partial m} = -t_u(1-\sigma(m))$,
i.e., the margin-gradient term in \eqref{eq:prop1_margin_template} is scaled by $t_u$.
Thus, the expected one-step margin change admits the first-order expansion
\begin{equation}
	\mathbb{E}[m^{+}-m]
	\;\ge\;
	\eta\,\mathbb{E}\!\left[t_u(1-\sigma(m))\|\nabla m\|_2^2\right]
	\;-\;O(\eta^2).
	\label{eq:prop2_weighted_margin}
\end{equation}
The first-order coefficient $\mathbb{E}\!\left[t_u(1-\sigma(m))\|\nabla m\|_2^2\right]$ scales linearly with $t_u$.
Since inactive users are assigned larger $t_u$, they obtain larger first-order margin gains per update,
increasing the probability that their positives outrank competing negatives and cross the top-$k$ boundary;
this tends to improve $P@k/R@k$ for inactive users.
For active users, although $t_u$ is smaller, their abundant interactions provide sufficient updates;
meanwhile, downweighting their gradients mitigates head-user dominance in the aggregated optimization,
reducing representation bias and potentially improving generalization. Our experiments in Section~4.4 further confirm these effects.

\subsection{Time Complexity}

The time complexity of PSP-NS mainly arises from three components: building the weighted bipartite graph, constructing positive sample pairs, and computing user weights. The first component includes reconstructing the interaction matrix using randomized SVD and reconstructing the bipartite graph via adaptive nearest neighbors. Randomized SVD in sparse matrix scenarios has an approximate complexity of $O(|E|q)$, where $|E|$ denotes the number of edges in the original bipartite graph and $q$ is the rank in SVD. 
Reconstructing the bipartite graph via adaptive nearest neighbors involves iterating and sorting operations, resulting in a complexity of $O(|U||P|\log|P|)$, where $|U|$ and $|P|$ represent the numbers of users and items in the original bipartite graph, respectively. 
The second component involves iterating over edges in the weighted graph to construct positive sample pairs based on edge confidence, with a complexity of $O(|\hat{E}|)$, where $|\hat{E}|$ denotes the number of edges in the weighted bipartite graph. The third component involves iterating over users to compute user weights based on activity, with a complexity of $O(|U|)$. Therefore, the overall time complexity of PSP-NS can be approximated as $O(|U||P|\log|P|)$.

\subsection{Discussion}

\textit{Position Analysis.} PSP-NS is a general negative sampling plugin because negative sampling depends on the quality of both positive and negative sample pairs, not negative samples alone. By constructing high-quality positive sample pairs, PSP-NS strengthens the supervision basis upon which negative sampling operates, since most samplers are conditioned on positives (e.g., sampling outside the positive set or selecting negatives by similarity to positives). Applicability study (Figure~4) further supports its generality and effectiveness.

\textit{Novelty Analysis.} The core novelty of PSP-NS lies in rethinking negative sampling from the perspective of enhancing positive supervision signals, rather than focusing on negative sample selection. In implicit CF, positive supervision signals are the primary source of information for learning user preferences, and negative samples are also derived from these signals. Therefore, by constructing high-quality positive sample pairs, PSP-NS explicitly guides the model to more accurately capture user preferences, resulting in substantial performance gains (see Table 2). To achieve this, SVD is employed to capture global interaction patterns and assist in constructing these positive pairs; however, it serves merely as a supporting tool, and the methodological contribution of PSP-NS fundamentally lies in its positive-signal-oriented design.

\begin{table}[t]  
	\centering
	\caption{The statistics of the datasets}
	\setlength{\tabcolsep}{4pt}           
	\begin{tabular}{ccccc}
		\toprule
		Dataset&Users&Items&Interactions&Density\\
		\midrule
		Pinterest& 55,187& 9,916&1,500,809&0.274\%\\
		Yelp& 29,061& 24,734&1,374,594&0.191\%\\
		Ml-1m& 6,040& 3,629&836,478&3.816\%\\
		Epinions& 11,496& 11,656&327,942&0.245\%\\
		\bottomrule
	\end{tabular}
\end{table}

\section{Experiments}

To demonstrate the effectiveness of PSP-NS, we conduct extensive experiments to answer: \textbf{RQ1}: How does PSP-NS perform compared to other baselines in implicit CF recommendation models? \textbf{RQ2}: How do parameters of PSP-NS affect the performance of recommendation tasks? \textbf{RQ3}: How well does PSP-NS improve negative sampling strategy via construction of positive sample pairs and calculation of user weights? \textbf{RQ4}: How does integrating PSP-NS into other implicit CF models (e.g., matrix factorization) perform? \textbf{RQ5}: How does integrating PSP-NS into other negative sampling methods perform?

\subsection{Experimental Settings}

\textbf{Datasets and Metrics.} We evaluate PSP-NS on four real-world datasets for implicit CF: Pinterest, Yelp, Ml-1m, and Epinions (statistics in Table~1). 
Pinterest~\cite{19} contains user–image interactions (e.g., collections/shares); Yelp~\cite{30} records user–merchants interactions (e.g., ratings/reviews); Ml-1m~\cite{31} provides MovieLens movie ratings; and Epinions~\cite{32} includes product ratings.
Following prior work, we adopt the same data splits: Yelp and Ml-1m use random partition with an 8:1:1 train/val/test ratio~\cite{30,31}; Pinterest follows SRNS~\cite{19} with 1,390,435/55,187/55,187 for train/test/val; and Epinions uses a 7:1:2 random split~\cite{32}.
We select widely used metrics in recommendations, Precision@k and Recall@k~\cite{10,19,23}, to evaluate the recommender performance, where k = [20,30].

\textbf{Baseline algorithms.} We compared PSP-NS with five negative-sample-focused baselines (RNS~\cite{56}, DNS~\cite{11}, MixGCF~\cite{39}, DNS(M,N)~\cite{48}, AHNS~\cite{49}) and five positive-sample-focused baselines (T-CE~\cite{51}, R-CE~\cite{51}, DeCA~\cite{52}, DCF~\cite{54}, PLD~\cite{55}). Unless otherwise specified, all baselines use the same hyperparameter settings as their original code implementations. 
To ensure a fair comparison, for all negative-sample-focused baselines, we adopt the user-interacted items in the original datasets as positives. For all positive-sample-focused baselines (including ours), we apply random negative sampling to obtain negatives.

\textbf{Hyperparameter Settings.} All comparative experiments are conducted on LightGCN~\cite{13}, with Xavier initialization~\cite{34}, embedding dimensions set to 64, employing Adam optimization~\cite{35}, the learning rate set at 0.001, the mini-batch size at 2048, the L2 regularization at 0.0001, and the model layers set to 3. The patience value for early stopping is set to 10. All comparative experiments are repeated 10 times with different random seeds, and we report the mean results.

\textbf{Remark.} All experiments accounted for potential data leakage: any user–item interactions present in the validation or test sets were excluded from the constructed PSP. 
Our code is available in the anonymous repository https://anony

\noindent mous.4open.science/r/PSP-NS-B105/.

\begin{table*}[t]
	\centering
	\caption{Overall Performance Comparison (\%). The best result is \textbf{bold}.}
	\resizebox{\linewidth}{29mm} {
		
		\begin{tabular}{c|ccc|ccc|ccc|ccc}
			\toprule
			\multirow{1}{*}{Dataset} & \multicolumn{3}{c|}{Pinterest} & \multicolumn{3}{c|}{Yelp} & \multicolumn{3}{c|}{Ml-1m} & \multicolumn{3}{c}{Epinions} \\
			\cmidrule(r){1-1} \cmidrule(r){2-4} \cmidrule(r){5-7} \cmidrule(r){8-10} \cmidrule(r){11-13}
			\multirow{1}{*}{Method} & R@20 & P@20 & R@30 & R@20 & P@20 & R@30 & R@20 & P@20 & R@30 & R@20 & P@20 & R@30 \\
			\midrule
			RNS      & 11.77 & 0.59 & 15.92 & 9.37  & 4.49 & 12.39 & 26.68 & 12.54 & 33.66 & 10.93 & 3.20 & 14.01 \\
			DNS      & 11.54 & 0.58 & 15.45 & 9.22  & 4.39 & 12.18 & 26.16 & 11.67 & 32.36 & 11.94 & 3.47 & 14.78 \\
			MixGCF   & 12.42 & 0.74 & 16.65 & 10.19 & 4.88 & 13.39 & 27.54 & 12.46 & 34.22 & 11.83 & 3.47 & 14.70 \\
			DNS(M,N) & 11.44 & 0.57 & 15.13 & 9.48  & 4.52 & 12.53 & 27.93 & 12.86 & 34.63 & 12.33 & 3.59 & 15.26 \\
			AHNS     & 11.71 & 0.59 & 15.86 & 9.57  & 4.57 & 12.68 & 27.47 & 12.67 & 34.12 & 11.66 & 3.40 & 14.70 \\
			\cmidrule(r){1-13}
			T-CE     & 10.13 & 0.51 & 13.76 & 7.19  & 3.55 & 9.67  & 23.56 & 11.93 & 29.91 & 11.17 & 3.20 & 14.11 \\
			R-CE     & 9.67  & 0.48 & 13.26 & 7.22  & 3.45 & 9.73  & 22.25 & 11.56 & 29.10 & 10.56 & 3.00 & 13.49 \\
			DeCA     & 10.69 & 0.53 & 14.75 & 8.47  & 4.12 & 11.36 & 25.32 & 12.24 & 32.09 & 11.33 & 3.29 & 14.60 \\
			DCF      & 10.73 & 0.54 & 14.68 & 8.17  & 3.99 & 10.91 & 24.20 & 12.27 & 31.20 & 11.15 & 3.23 & 14.30 \\
			PLD      & 10.93 & 0.55 & 14.78 & 8.68  & 4.11 & 11.57 & 26.22 & 12.36 & 32.52 & 11.80 & 3.43 & 14.99 \\
			\cmidrule(r){1-13}
			PSP-NS   & \textbf{18.03} & \textbf{0.90} & \textbf{21.48}
			& \textbf{15.01} & \textbf{8.25} & \textbf{17.69}
			& \textbf{39.56} & \textbf{21.31} & \textbf{45.89}
			& \textbf{13.01} & \textbf{4.50} & \textbf{15.38} \\
			\bottomrule
	\end{tabular} }
\end{table*}

\begin{figure}[t]
	
	\includegraphics[width=\linewidth]{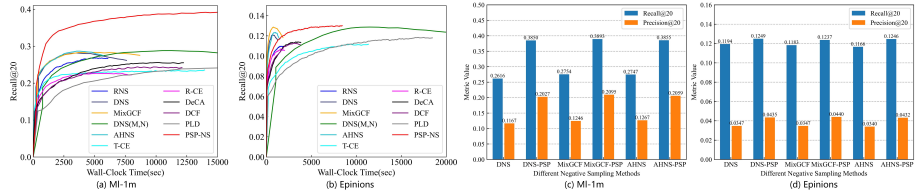}
	
\caption{(a)-(b) Recall@20 vs. wall-clock time (s). (c)-(d) Performance of integrating PSP-NS with different negative sampling methods.}
	
\end{figure}

%

\begin{figure*}[t]
	\includegraphics[width=\textwidth]{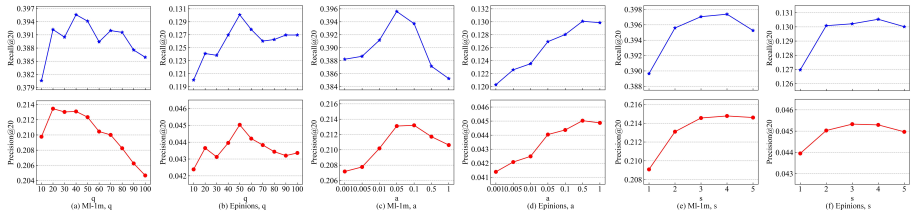}
	
	\caption{The impact of $q$, $a$, and $s$ on Recall@20 and Precision@20.}
	
\end{figure*}

\subsection{Performance Comparison (RQ1)}

Table~1 reports the comparative results, where PSP-NS achieves significant improvements across all metrics and datasets over strong baselines. This stems from (i) constructing high-quality positive sample pairs by combining global patterns and local signals, which both suppresses noisy signal and strengths positive supervision signal, and (ii) activity-aware user weighting that focus on inactive users during model training and mitigates activity bias.
Additionally, Denoising recommendation methods often underperform
because they rely on dataset-specific observations, limiting their applicability, and use statistical features from model training while ignoring contextual information, weakening positive signals and hindering the accuracy of learning user preferences.

Besides effectiveness, we evaluate the efficiency of PSP-NS. We measured the time required to construct positive sample pairs, recording 35.14s, 12.90s, 3.03s, and 1.89s on the Pinterest, Yelp, ML-1M, and Epinions datasets, respectively. More importantly,  PSP-NS is a one-time offline process before model training. This demonstrates that PSP-NS can efficiently generate high-quality positive sample pairs. Figure~2(a)-(b) shows that PSP-NS achieves higher performance in less time and with greater stability, demonstrating its superior efficiency.


To assess robustness, we run PSP-NS with 10 random seeds on all datasets. The variations of Recall@20/Precision@20 are within 0.78\%/1.11\% (Pinterest), 0.67\%/0.48\% (Yelp), 1.14\%/0.42\% (ML-1M), and 3.15\%/2.67\% (Epinions), indicating that PSP-NS produces highly stable performance across random seeds.

\subsection{Parameter Analysis (RQ2)}


We conduct parameter analyses for the retained rank $q$, the edge weight $s$ in $\hat G$, and the user-weight sensitivity $a$, to show their effects on performance. 
Depending on the training set scale, $q$ ranges from \{50, 100, .., 500\} on Pinterest and Yelp; and from \{10, 20, …, 100\} on Epinions and Ml-1m. 
Since larger $s$ increases the scale of positive sample pairs and training cost, we select $s\in\{1,2,3,4,5\}$. We vary $a\in\{0.001,0.005,0.01,0.05,0.1,0.5\}$. When analyzing one parameter, the other parameters remain fixed.

Figure 3(a)-3(b) show the impact of $q$. It can be observed that Recall@20 and Precision@20 increase with $q$ at first and then decrease. This is because when $q$ is small, the SVD-reconstructed interaction matrix may lose information reflecting users' true preferences, which negatively impacts model performance. When $q$ is large, the SVD-reconstructed user-item interaction matrix may capture noise from the interaction data, reducing the accuracy of recommendations.

Figure~3(c)–3(d) show that Recall@20 and Precision@20 first increase with $a$ and then decrease. When $a$ is small, user weights are highly skewed, so gradients are dominated by a few inactive users, biasing optimization and hurting generalization. 
When $a$ is large, weights become nearly uniform, leaving inactive users with insufficient training signal, which also degrades performance.

Figures~3(e)–3(f) show the effect of $s$. 
Larger $s$ replicates high-confidence positives more and thus increases their share in the constructed PSP. 
Therefore, analyzing $s$ essentially explores how the positive sample distribution constructed by PSP-NS affects model performance. 
As shown in Figures~3(e)–3(f), Recall@20 and Precision@20 generally increase first and then decrease as $s$ grows. 
This is because moderate $s$ strengthens reliable positive supervision and improves preference learning, while overly large $s$ overemphasizes a subset of positives, leading to overfitting, which hurts performance.

\begin{table}[t] 
	\centering 
	\caption{Performance comparison of different variants (\%).}
		\setlength{\tabcolsep}{4pt}           
			\begin{tabular}{c|cccc|cccc}
				\toprule
				\multicolumn{1}{c|}{Dataset} & \multicolumn{4}{c|}{Ml-1m} & \multicolumn{4}{c}{Epinions} \\ 
				\cmidrule(lr){1-1} \cmidrule(lr){2-5} \cmidrule(lr){6-9}
				\multirow{1}{*}{Method} & R@20 & P@20 & R@30 & P@30 & R@20 & P@20 & R@30 & P@30 \\ 
				\midrule
				1-hop              & 26.68 & 12.54 & 33.66 & 10.95 & 10.93 & 3.20 & 14.01 & 2.75  \\
				1-hop*2            & 26.49 & 12.38 & 33.34 & 10.84 & 10.81 & 3.16 & 13.74 & 2.70  \\
				SVD-hop             & 34.19 & 19.51 & 39.30 & 15.68 & 9.61 & 3.52 & 11.14 & 2.69  \\
				W-hop    & 37.91 & 20.52 & 44.07 & 16.73 & 12.36 & 4.27 & 14.52 & 3.31  \\
				W-hop-lw        & 38.25 & 20.74 & 44.55 & 16.94 & 12.46 & 4.30 & 14.67 & 3.35 \\
				W-ew        & 38.40 & 20.92 & 44.70 & 17.10 & 12.62 & 4.38 & 15.02 & 3.42  \\ \cmidrule(r){1-9} 
				ISW                & 39.29 & 21.16 & 45.53 & 17.30 & 12.77 & 4.44 & 15.33 & 3.48  \\
				EDW                & 39.22 & 21.12 & 45.44 & 17.22 & 12.66 & 4.39 & 15.11 & 3.43  \\
				CRW                & 39.34 & \textbf{21.31} & 45.55 & 17.40 & 12.73 & 4.40 & 15.15 & 3.44  \\
				\cmidrule(r){1-9} 
				PSP-NS             & \textbf{39.56} & \textbf{21.31} & \textbf{45.89} & \textbf{17.41} & \textbf{13.01} & \textbf{4.50} & \textbf{15.38} & \textbf{3.50} \\ 
				\bottomrule
			\end{tabular}

\end{table}

\subsection{Ablation Study (RQ3)}


To evaluate how PSP construction and user weighting improve negative sampling, we compare PSP-NS with eight variants. 
(i) PSP construction variants. 
1-hop builds PSP from 1-hop neighbors in the original graph $G$, and 1-hop*2 repeats it twice. 
SVD-hop uses 1-hop neighbors in the SVD-reconstructed graph $G_{\mathrm{SVD}}$, and W-hop applies the same strategy on the weighted bipartite graph $\hat G$. 
W-hop-lw uses the same PSP construction as W-hop and weights the loss of each triplet by the positive-edge weight.
W-ew constructs PSP directly based on edge weights in $\hat G$ via replication-based reweighting.
None of these uses activity-aware user weights.
(ii) User-weighting variants. 
Fixing PSP construction to W-ew, ISW, EDW, and CRW apply inverse square-root, exponential-decay, and capped-reciprocal weighting for user activity, respectively. For fairness, 
we introduce the same sensitivity-control factor into these weighting functions and tune it over the same search range.

Table 4 shows the results of different variants. We observe the following phenomena: 
[1] W-hop exceeds SVD-hop and 1-hop, demonstrating the benefit of combining global patterns with local signals for PSP construction.
[2] W-ew $>$ W-hop-lw $>$ W-hop indicates that exploiting edge-weight confidence is beneficial, and replication-based reweighting is particularly effective under negative sampling by providing richer supervision than loss-only weighting. 
[3] W-ew also surpasses 1-hop*2 and W-hop, suggesting that gains come from higher-quality PSP rather than simple duplication.
[4] PSP-NS, ISW, EDW, and CRW consistently improve over W-ew, confirming the importance of activity-aware user weighting.
[5] PSP-NS performs best among user-weighting variants because its logarithmic weighting sharply boosts the learning signal of inactive users while suppressing that of highly active users in a smooth and stable manner. In contrast, the other weighting functions cannot achieve both effects simultaneously.

\begin{table}[t]
	\centering
	\caption{Performance comparison of W-ew and PSP-NS for inactive users and other users, where PSP-NS(i) indicates PSP-NS on inactive users and W-ew(o) indicates W-ew on other users (\%).}
			\setlength{\tabcolsep}{4pt}           

			\begin{tabular}{c|cc|cc|cc|cc}
				\toprule
				\multirow{1}{*}{Method} & \multicolumn{2}{c|}{W-ew(i)} & \multicolumn{2}{c|}{PSP-NS(i)} & \multicolumn{2}{c|}{W-ew(o)} & \multicolumn{2}{c}{PSP-NS(o)} \\ \midrule                                                                                                                   
				\multirow{1}{*}{Dataset} & \multicolumn{1}{c}{R@20} & \multicolumn{1}{c|}{P@20} & \multicolumn{1}{c}{R@20} & \multicolumn{1}{c|}{P@20} & \multicolumn{1}{c}{R@20} & \multicolumn{1}{c|}{P@20} & \multicolumn{1}{c}{R@20} & \multicolumn{1}{c}{P@20} \\ \midrule
				Yelp       & 13.88 & 3.27 & 14.42 & 3.39 & 14.71 & 8.34 & 14.95 & 8.37 \\ 
				Ml-1m      & 42.18 & 3.89 & 42.97 & 3.98 & 37.69 & 24.10 & 38.66 & 24.75 \\ 
				Pinterest  & 48.60 & 2.43 & 50.60 & 2.53 & 17.14 & 0.86 & 17.37 & 0.87 \\ 
				Epinions   & 11.97 & 1.54 & 12.62 & 1.65 & 12.58 & 4.61 & 12.85 & 4.73 \\ \bottomrule
			\end{tabular}

\end{table}

Table~5 reports the effect of activity-aware user weighting on inactive vs. other users. 
We define inactive users as the 1,000 users with the fewest interactions in the training set. 
Activity-aware user weighting consistently improves performance for inactive users across all datasets, and also benefits the other users. 
We attribute this to down-weighting highly active users’ gradient contributions (see Proposition~2), which reduces overfitting to active users and yields more balanced, generalizable representations.

\begin{table}[t]
	\centering
	\caption{Performance comparison integrated with MF (\%).}
	\setlength{\tabcolsep}{4pt}           
	\begin{tabular}{c|cccc|cccc}
		\toprule
		\multicolumn{1}{c|}{Dataset} & \multicolumn{4}{c|}{Ml-1m} & \multicolumn{4}{c}{Epinions} \\ 
		\cmidrule(lr){1-1} \cmidrule(lr){2-5} \cmidrule(lr){6-9}
		\multirow{1}{*}{Method} & R@20 & P@20 & R@30 & P@30 & R@20 & P@20 & R@30 & P@30 \\ 
		\midrule
		RNS            & 23.63 & 11.05 & 30.18 & 9.74  & 9.18 & 2.65 & 11.77 & 2.29  \\
		DNS            & 20.58 & 9.46 & 26.09 & 8.26  & 8.47 & 2.47 & 10.55 & 2.09  \\
		DNS(M,N)       & 22.97 & 10.86 & 29.15 & 9.48  & 9.45 & 2.75 & 11.84 & 2.31  \\
		AHNS           & 23.42 & 11.27 & 29.53 & 9.86  & 9.58 & 2.84 & 12.21 & 2.43  \\				\cmidrule(r){1-9} 
		T-CE           & 22.93 & 10.80 & 29.87 & 9.74 & 8.24 & 2.35 & 10.51 & 2.01  \\ 
		R-CE           & 22.12 & 10.12 & 28.59 & 9.08  & 8.46 & 2.45 & 11.07 & 2.14\\
		DeCA           & 23.11 & 11.63 & 29.28 & 10.12 & 8.79 & 2.58 & 11.18 & 2.21\\ 
		DCF            & 23.90 & 11.17 & 30.75 & 9.92  & 8.80 & 2.51 & 11.38 & 2.19  \\
		PLD            & 25.59 & 12.20 & 32.36 & 10.66 & 8.44 & 2.56 & 10.61 & 2.18\\ 	\cmidrule(r){1-9} 
		PSP-NS         & \textbf{36.57} & \textbf{19.32} & \textbf{42.56} & \textbf{15.81} & \textbf{11.46} & \textbf{3.90} & \textbf{13.89} & \textbf{3.11} \\ 
		\bottomrule
	\end{tabular}
	
\end{table}

\subsection{Applicability Study (RQ4, RQ5)}

In this section, we integrate PSP-NS with other implicit CF recommendation models and negative sampling methods to evaluate its applicability.

Table~6 reports results on an MF-based recommender~\cite{44}. We exclude MixGCF since it is tailored to GNN aggregation and is not applicable to MF. 
PSP-NS consistently outperforms the baselines, demonstrating its effectiveness and plug-and-play transferability to different implicit CF models.

We integrate PSP-NS with three popular negative samplers (DNS, MixGCF, and AHNS), yielding DNS-PSP, MixGCF-PSP, and AHNS-PSP. 
For each, PSP-NS provides constructed PSP and user weights, while the base method supplies negatives to form the final negative sampling strategy that is used to train the model. 
As shown in Figure~4, PSP-NS consistently boosts their performance, demonstrating its effectiveness and broad applicability.


\section{Conclusion}
\label{conclusion}

In this paper, we revisit negative sampling in implicit CF from a new perspective that constructs high-quality positive sample pairs and propose PSP-NS, a simple and effective negative sampling plugin with two components. 
The first component builds a weighted user–item bipartite graph by capturing both global interaction patterns and local interaction signals, and then generates high-quality positive sample pairs via edge-weight-guided replication-based reweighting to emphasize high-confidence interactions, enabling the model to learn users’ true preferences more accurately.
The second component adaptively calculates user weights based on user activity level, allowing the model to focus more on inactive users during training, thus reducing user activity bias. 
Theoretical insights and extensive experiments show that PSP-NS consistently improves existing CF models trained with negative sampling.
Its strong experimental results also indicate that focusing on constructing high-quality positive sample pairs within negative sampling methods can further improve the performance of implicit CF recommendation models, providing a promising new direction for future research.

%
%
%
 \bibliographystyle{splncs04}
 \bibliography{citiation}
%
%
%
%
%
\end{document}